\documentstyle[12pt,epsf]{article}

\begin{document}

\title{
Do Instantons and Strings Cluster when the Number of Colors is Large?
}

\author{G.~W.~Carter and E.~V.~Shuryak \\
\small{\em Department of Physics and Astronomy,} \\
\small{\em State University of New York, Stony Brook, NY 11794-3800}
}
\date{\today}
\maketitle

\begin{abstract}
We consider the $N_c\rightarrow\infty$ limit of QCD using a toy model in which 
instantons exchange color-singlet scalar fields which do not self-interact.
Our main observation is that collective attraction leads the formation of 
large clusters containing $O(N_c)$ nonperturbative objects.
We further show that this clustering of instantons is limited due to a 
non-trivial repulsion inherent in the ADHM multi-instanton solution.
As a result the vacuum is very different from that at
low $N_c$, notably being more inhomogeneous, in ways which will affect
chiral symmetry breaking of light quarks. 
We also briefly discuss a similar phenomenon for color strings
in baryons made of medium-mass (charm-like) quarks. 
\end{abstract}

When localized in sufficiently large numbers, even weakly interacting bosons
can bind together.
Although, for example, electroweak interactions between $W$, $Z$, and Higgs
bosons are of order $\alpha_w \sim 1/100$, a hundred or more may collectively
bind and form a sphaleron.
In this paper we consider similar phenomena with nonperturbative QCD objects, 
concentrating on instantons but eventually turning to strings as well.

In QCD, the instanton liquid model \cite{Shu_82,DP}
has been consistently successful in describing
the physics of light hadrons (see Ref.~\cite{SS_98} for a review). 
Of central importance is its description of phenomena related to chiral
symmetry breaking, confirmed in detail by recent lattice studies of the 
lowest Dirac eigenvalues \cite{Negele,DH00}.  
The primary parameters of this model are the number density of instantons
(plus anti-instantons) and their average size, determined long ago 
from phenomenology to be $n \approx 1$~fm$^{-4}$ and 
$\bar\rho\approx 1/3$~fm, respectively \cite{Shu_82}.
With these numbers one obtains an outstanding quantitative description of 
hadronic correlation functions; for example, the most accurately known 
vector and axial correlators (from  $\tau$ decays) 
are reproduced at all distances literally within the (rather small) 
experimental error bars \cite{SS_00}.

At the same time, this picture of the QCD vacuum remains incomplete.  
The instanton ensemble does not explain confinement effects. 
Furthermore, the apparent suppression of large-size instantons remains 
unexplained dynamically\footnote{ 
For a recent attempt to relate the small size of instantons with a small radius
of QCD strings see \cite{Shu_99}.}.

Other open questions are related to the behavior of 
the instanton ensemble in the large $N_c$ limit. 
Long ago, Witten emphasized that small-size instantons should be 
$exponentially$ suppressed \cite{Wit_79} but, on
the other hand, he recently found that the topological susceptibility
in large $N_c$ gluodynamics is merely $power$ suppressed \cite{Wit_98},
as $\chi\sim \Lambda^4 O(N_c^0)$. 
In general, there are many possible scenarios in which instantons of size 
$\rho\sim 1/\Lambda$ can satisfy both conditions.   
The first steps towards a numerical resolution of this issue were recently 
made in Ref.~\cite{LT}, in which instantons were studied for $N_c=2-5$. 
Although these authors found a suppression of small-sized instantons, 
those of greater sizes are enhanced, suggesting that the total topological 
succeptibility is finite at large $N_c$.

In this letter we begin by noticing that in the $N_c=3$ instanton vacuum 
the gauge fields are distributed inhomogeneously such that the field strength
is concentrated in small regions of space-time, the diluteness parameter
being $n\bar\rho^4\sim (1/3)^4$.
The gluonic correlators consequently decrease much more rapidly with distance 
than those of mesons containing light quarks, giving the ``glueball''
spectrum a mass scale larger than that of the light mesons.

As $N_c\rightarrow\infty$, we will suggest that the instanton ensemble
becomes even {\em more inhomogeneous} than the two and three color cases 
studied to date.
This idea apparently contradicts the earlier view that in the large-$N_c$ limit 
the so-called ``master field'' must be simple and homogeneous \cite{EK}.
And yet this picture does agree with recent progress regarding the
$N_c\rightarrow\infty$ limit of ${\cal N}=4$ supersymmetric Yang-Mills 
(SUSY YM) theory. 
In that case the master field is an {\em instanton cluster} in which all 
instantons share a common location and size \cite{hkm}. 
This is in perfect agreement with Maldacena's conjecture of the AdS/CFT 
correspondence \cite{mald}, 
where the Anti-deSitter 5-d space is manifested simply as the
instanton measure of collective coordinates, $d^4zd\rho/\rho^5$.

Despite the very profound differences between ${\cal N}=4$ SUSY
YM theory and QCD-like theories -- the former is a conformal theory without 
dimensional parameters, and clustering results from the exchange 
of adjoint fermions not present in QCD --
we argue below that this phenomenon is in fact rather generic.
Indeed, similar clustering of instantons should also occur in QCD at
large $N_c$.

Our primary argument for this conjecture is based on the idea that,
although they do not interact via direct color forces, 
gluonic objects of completely different colors are generically attracted
through effective interactions.
We will specifically describe instantons exchanging a colorless scalar
field.
Although the corresponding coupling becomes weak at large $N_c$, this 
weakness is overcome by the large ($O(N_c^2)$) number of pairs and 
clustering ensues\footnote{
Note that this attraction is similar to that between nucleons at large
distances.  However, unlike instantons or strings, they are fermions and
as such have a strong repulsive core which leads to a saturation density in
their clusters (nuclei).
}.

The toy model we use couples the gauge fields to a color-singlet scalar field,
which can be interpreted as the $0^{++}$ glueball.
This scalar naturally interacts with {\em any} gluonic configurations --
instantons, flux tubes, etc.
Not only is this the lightest dynamical state in pure glue theory, it
is also has been found to be extraordinarily small, with a size of about 
0.2~fm \cite{fl}.  
For this reason it may be reasonable to consider it as an elementary 
dynamical field, analogous to pion-mediated dynamics in nuclei.
The action we have in mind is an Euclidean one with the simplest form
encoding our degrees of freedom:
\begin{equation}
S = \int d^4x \, \Big[ 
\frac{1}{4g^2} G_{\mu\nu}^a(x) G_{\mu\nu}^a(x)
+ \frac{1}{2} \partial_\mu\phi(x) \partial_\mu\phi(x)
+ \lambda \phi(x) G_{\mu\nu}^a(x) G_{\mu\nu}^a(x) \Big].
\label{action}
\end{equation}
Here the gauge field strength tensor is
$G_{\mu\nu}^a = \partial_\mu A_\nu^a - \partial_\nu A_\mu^a +
f^{abc}A_\mu^a A_\nu^a$, the
parameter $\lambda$ is a coupling constant with the dimension of
length, and as usual the factor $g$ has been absorbed in the gauge field.
While one could naturally include a potential for the field 
$\phi$ we see no motivation to do so in this case.
Minimizing Eq.~(\ref{action}) generates the equations of motion:
 \begin{eqnarray}
 \left(\frac{1}{4g^2}+\lambda\phi(x)\right)D_\mu G_{\mu\nu}^a(x) &=& 0 \,,
 \label{a_eom}
 \\
 \partial^2\phi(x) - \lambda G_{\mu\nu}^a(x)G_{\mu\nu}^a(x) 
 &=& 0 \,.
 \label{phi_eom}
 \end{eqnarray}
The self-dual instanton solution, unchanged by the presence of $\phi$, is
\begin{equation}
A_\mu^a(x) = \frac{2\eta_{\mu\nu}^a x_\nu}{(x-z)^2 + \rho^2} \,,
\end{equation}
where $\eta_{\mu\nu}^a$ is the usual 't Hooft symbol, with $\rho$ and $z$ the
the instanton size and position.

The parameter $\lambda$ can be determined by the scalar
glueball mass, $m_\phi \simeq 1.5$ GeV.
After squaring the interaction term and Fourier transforming the non-local
result, the leading contribution to the glueball self-energy is
\begin{equation}
\Pi(q^2) = \left(32\pi^2\right)^2\lambda^2 \int\! d\rho\, \nu(\rho)
\,(q\rho)^4 K_2(q\rho)^2 \,,
\end{equation}
where $K_2(x)$ is a modified Bessel function.
From the static limit the mass is obtained,
\begin{equation}
m_\phi^2 = \Pi(0) = (32\pi^2)^2 \lambda^2 n \,.
\end{equation}
With this one finds $\lambda = 0.024\, {\rm fm}$ in order to fit a glueball
mass of 1.5 GeV.
The glueball mass is 
assumed to be stable with respect to the number of colors
\cite{nsvz81}, whereas the instanton density is assumed to scale
as $n \sim N_c$ according to Ref.~\cite{DP}.
Thus we must have a decreasing coupling constant
\begin{equation}
\lambda = \frac{\lambda_0}{\sqrt{N_c}} \, ,
\label{lam_sc}
\end{equation}
where $\lambda_0$ may be fixed by the previous case of $N_c=3$.

The attraction between any pair of pseudo-particles is quantified by the
change in the action, parametrized as usual by the instanton sizes and the
separation between their centers, $R$.
For two isolated instantons, this is
\begin{equation}
\Delta S_{pair}(R,\rho_1,\rho_2) = -\frac{\lambda^2}{2} \int d^4x d^4y
\left[G_{\mu\nu}(x-R,\rho_1)\right]^2
D(x-y) \left[G_{\mu\nu}(y,\rho_1)\right]^2 \,.
\label{exch_DS}
\end{equation}
where the scalar propagator in Euclidean space is
\begin{equation}
D(x) = \frac{m}{4\pi^2 |x|} K_1(m|x|) \,,
\end{equation}
and the sources are the field strength tensor evaluated over an instanton
action distribution.
This will describe the extent of interactions between two instantons in
completely distinct SU(2) color subgroups, as well as the interaction
between an instanton and anti-instanton.

According to Eqs.~(\ref{lam_sc}) and (\ref{exch_DS}), the attractive force 
between any pair of pseudo-particles will fall as $1/N_c$ due to the reduced
coupling constant.
Yet this force remains relevant, and in fact becomes dominant, when one
considers a cluster of instantons with a number of constituents on the order
of the number of colors.
Specifically, the total change in the action from such a cluster is
parametrically large:
\begin{equation} 
\Delta S_{clust} = \sum_{i>j}\, \Delta S(R,\rho_i,\rho_j)
\sim N_c^2 \overline{\Delta S_{pair}} \sim N_c \,.
\end{equation}
Thus it will overcome the entropy of random placements, so that
the large-$N_c$ ensemble will be one of 
strongly bound instanton clusters with $\sim N_c$ constituents.
This force will be strongest between instantons which share a common size,
and in this way the size distribution will be sharply narrowed.

Can such a clustering process continue indefinitely, until
all of them collapse into the same mega-cluster?
The answer is negative for two reasons. 
The first is that instantons and anti-instantons in the same SU(2) subgroup 
would obviously annihilate one another.
The second, less trivial, effect which would limit clustering is the rather 
interesting non-linear deformation of the two-instanton solution. 
Although this is, strictly speaking, only derived for instanton pairs in the 
same color subgroup, we think it will generally lead to a kind of
shell model situation (albeit without fermions) in which instantons and 
anti-instantons tend to occupy only $non-overlapping$ SU(2) subgroups.
We now consider both effects in detail.

Any group of pseudo-particles will likely include as many anti-instantons
as instantons, for the scalar exchange knows no difference\footnote{
Consequentially, in this ensemble the instantons and
anti-instantons of different colors are statistically
independent and therefore its topological susceptibility
is {\em not suppressed} relative to an unclustered ensemble, 
$\chi\sim N_c \Lambda^4$. 
A more refined model with more complicated interactions is therefore needed 
to provide for Witten's suppression, $\chi\sim\Lambda^4$, such as one that
restricts the topological charge of each cluster to $Q= O(1)$.
}. 
The inevitable annihilation between instantons and anti-instantons within a 
given cluster will leave some fraction of the original
number in a cluster of instantons or anti-instantons alone\footnote{
In the calculations which follow all statements about instanton clusters
naturally apply to those made of anti-instantons.}.
Statistical fluctuations in $N_I - N_{\bar I}$ will however guarantee that
these clusters persist, with a number of constituents still on the order of
$N_c$.

Instantons in the same SU(2) subgroup may be analyzed using an
exact multi-instanton solutions known as the ADHM construction \cite{ADHM}.
We find that this leads to an effective repulsion for two reasons.

First, when two such pseudo-particles converge, their distinct
peaks in the action density are deformed continuously into a toroidal
cloud in two of the four spatial directions \cite{gkvb}.
There is a minimal separation, $R_{min}=\sqrt{\rho_1\rho_2/2}$, after
which a certain coordinate transformation reveals re-separation into a
direction orthogonal to the original $R_\mu$.
Although the classical action remains constant, this modifies
the non-trivial measure of the two-instanton coordinate space
at the quantum level (in the pre-exponent of the partition
function).
So although the manifold is smooth, we deduce that a coordinate discontinuity 
must exist to ensure that the same configuration is not counted twice
(an observation, to the best of our knowledege, not made in literature before).
This minimal separation is an effective hard-core repulsion between 
instantons of the same color projection.

The second effective repulsion arises from interactions between different
color subgroups.
Note that the action density becomes less concentrated as two instantons 
in the same subgroup are deformed into a toroidal-like configuaration.
Consider now the interaction between this toroidal configuration 
and a separate cluster of instantons, each of a different adjoint color
and hence having unmodfied shapes. 
The following calculation reveals that the total attraction in this case is
less than that of a cluster interacting with two color-independent, 
unmodified instantons.
Indeed, instead of Eq.~(\ref{exch_DS}) we now would have
\begin{equation}
\Delta S_{pair}(R,\rho_1,\rho_2) = -\frac{\lambda^2}{2} \int d^4x d^4y
\left[G_{\mu\nu}(x-R,\rho_1)\right]^2
D(x-y) \frac{1}{2}s(y,\rho_2,\rho_2) \,,
\end{equation}
where
\begin{equation}
\frac{1}{4} s(x,\rho_1,\rho_2) = -\frac{1}{2} \partial^2\partial^2
{\rm ln}\,{\rm det}\left[\Delta(x)^\dagger\Delta(x)\right]
\end{equation}
is the action density for the ADHM solution.

The matrix $\Delta(x)$ is a rather complicated function and we refer the 
interested reader to Ref.~\cite{gkvb} for a thorough analysis.
Since we are only interested in quantitative scalar exchanges between an 
instanton and such a solution, we consider two instantons
in the same SU(2) subgroup and with a color group angle of $\pi/2$ between
the two (the case of maximal deformation).
In that case, we have
\begin{eqnarray}
{\rm det}\left[\Delta(x)^\dagger\Delta(x)\right] &=&
\left[\left(x-R\right)^2 + \rho_1^2 + \frac{(\rho_1\rho_2)^2}{R^2}\right]
\nonumber\\
&& \hspace{-2cm}
\times\left[x^2 + \rho_2^2 + \frac{(\rho_1\rho_2)^2}{R^2}\right]
- \frac{(2\rho_1\rho_2)^2}{R^2}x_1^2\,.
\end{eqnarray}
Here we have chosen $R_\mu = (R,0,0,0)$ and color group elements 
$\sigma_0$ and $\sigma_1$ for the two pseudo-particles, which leads to a
torus in the plane of coordinates $x_0$ and $x_1$.

\begin{figure}[t]
  \epsfxsize=4.0in
  \centerline{\epsffile{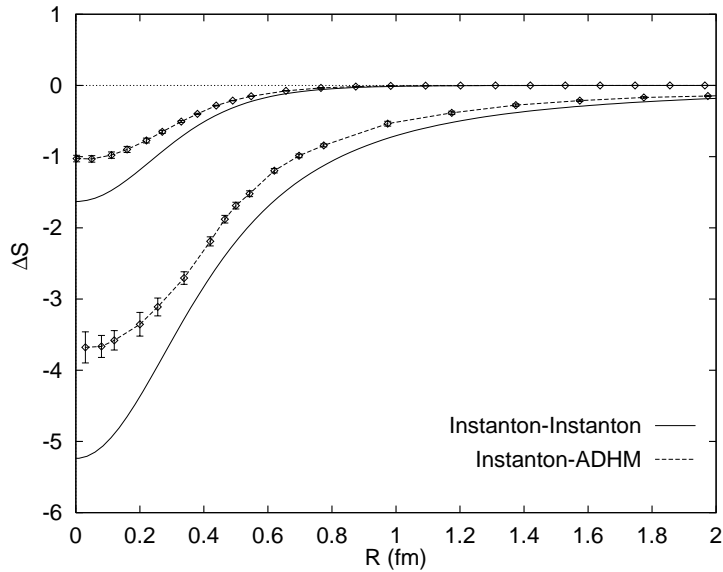}}
  \caption{
Interaction strength of scalar exchange between either 
an instanton-instanton pair or instanton-ADHM pair as a function of separation,
for $m_\phi = 1500$ MeV (upper lines) and $m_\phi=0$ (lower lines)
($\rho_1=\rho_2=1/3$ fm for all).
  }
\label{sc_exch_fig}
\end{figure}

With this specification we have computed via Monte-Carlo the ``binding
action'' generated by scalar exchange between an instanton and half an
ADHM pair, which is compared to that for two more prosaic instantons in
Fig.~\ref{sc_exch_fig}.
From this we can see that the cluster formation will prefer filling each
SU(2) subgroup once to allow for maximal attraction.
This generates a medium of isolated clusters with the order of $N_c$
individual instantons.

We now depart from pure glue theory and consider the effects of light
fermions on the clustering process.
Specifically, we discuss QCD-like fermions with fundamental color charges 
rather than the gluino-like adjoint fermions present in supersymmetric 
theories\footnote{
The difference is crucial, since the former will have $O(N_f)$ and the latter
$O(N_c)$ fermionic zero modes.}.

The clustering of instantons as described  effectively reduces the 
instanton density, $n$, by a factor of $N_c$.
This will have severe consequences for the spontaneous breaking of chiral
symmetry.
In a background of unclustered instantons, the chiral condensate and 
quark effective mass scale as \cite{DP}:
\begin{equation}
\langle\bar\psi\psi\rangle \sim N_c \,,\quad
M_q \sim N_c^0 \,.
\end{equation}
In a clustered environment we will instead have
\begin{equation}
\langle\bar\psi\psi\rangle \sim \sqrt{N_c} \,,\quad
M_q \sim \frac{1}{\sqrt{N_c}} \,.
\end{equation}
Spontaneous chiral symmetry breaking is thus much stronger with unclustered 
instantons,
and a large number of quark flavors might keep the system in a phase of 
strongly broken chiral symmetry rather than the clustered medium.
To estimate how many flavors would be necessary, we note that the fermion
determinant built of zero-mode overlap matrices would raise the action,
possibly overcoming the reduction from scalar exchange (\ref{exch_DS}) which
will appear in the exponent of the partition function.
For clustering to persist we must have
\begin{equation}
{\rm det}\left({\cal T}\right)^{2N_f} e^{-N_c\Delta S_{pair}} 
\equiv e^{-\Delta S_{eff}} > 1 \,,
\end{equation}
where ${\cal T}$ is the matrix of overlaps.
To extract the $N_c$ dependence from the determinant, we note that the
overlap matrices involve two zero-mode propagators and hence 
\begin{equation}
\ln\,{\rm det}\left({\cal T}\right) \sim \ln\left(\frac{R^3}{R_0^3}\right)
\sim \frac{3}{4}\ln\left(\frac{n_0}{n}\right)
\sim \frac{3}{4}\ln N_c \,.
\end{equation}
The total effective change in the action is thus
\begin{equation}
\Delta S_{eff} = \frac{3}{2}N_f\ln N_c + N_c\Delta S_{pair} \,.
\end{equation}
This becomes positive -- and clustering becomes unstable -- for
\begin{eqnarray}
N_f &>&  - \frac{2}{3} \Delta S_{pair} \, \frac{N_c}{\ln N_c} 
\nonumber\\
&{\buildrel > \over \sim}& \frac{2N_c}{\ln N_c} \,,
\end{eqnarray}
where a typical value for $\Delta S_{pair}$ has been inserted.
Thus we see that the clustering phenomena will be inhibited by a rather
small $N_f/N_c$ value.

As already mentioned, the arguments presented above would likely apply not 
only to instantons but to any configuration of color fields.
Color flux tubes (QCD strings) in particular
are highly localized color field distributions, in two rather than four 
dimensions. 
Assuming flux tubes of different colors can also interact via colorless 
glueball exchange, the strings will cluster in the large-$N_c$ limit in 
a similar fashion.
Profound consequences for hadronic structure would follow.
 
Consider for example a baryon made of medium-mass (charm-like) 
quarks\footnote{
Hadrons made of light quarks have different features and baryons made
of them are skyrmion-like, whereas those made of very heavy ($b$-like) quarks 
are dominated by perturbative Coulomb forces. 
Only charmonium spectroscopy is described well by a linear potential 
and flux tubes.
}.  
It has $N_c$ flux tubes originating from each quark and ending
in a central ``junction''.  The standard picture would be that the quark
motion is described by a 3-d string potential, 
$V(x)\sim |\vec{x}|$, around the junction. However 
if the flux tubes spontaneously cluster along one direction, as we suggest, 
quark motion will be constrained to that direction and the potential becomes
one dimensional.
If so, the large classical deformation of such baryons leads to
large momentum of inertia and an unusually soft rotational band.

In addition to the colorless attractive forces described here in our toy 
model, clusters could also form due to confinement effects,
if confinement is indeed due to a dual superconductor
and its universal U(1) dual Higgs mechanism \cite{dualsuper}.
Because the Higgs VEV must vanish at the center of {\em any} topological
object, including both instantons \cite{Shu_99} and flux tubes,
it will be energetically favorable to have these objects at a minimal
number of locations.
The Higgs field has been shown numerically to be rather robust for small
$N_c$, restricting the color flux tubes to small radii (and perhaps instantons
to small sizes \cite{Shu_99}), and so it easy to imagine its effects playing
a major role in the large-$N_c$ limit.

How might our predictions be tested?
Perhaps most directly by lattice simulations.
Color groups of currently-simulated SU(4) \cite{Ohta} and higher contain
orthogonal SU(2) subgroups, allowing for computation of the correlation 
strengths between instantons and/or color strings.
The number of colors could also be increased to computational limits,
and if, as we expect, the nonperturbative objects cluster, many observables
would show very different qualitative behavior.
Lattice calculations along these lines would provide the most reliable test
of vacuum configurations in the large-$N_c$ limit.

{\em Acknowledgement:}
This work is partially supported by the US-DOE grant No. DE-FG02-88ER40388.

\end{document}